# Broadband Parametric Impedance Matching for Small Antennas Using the Bode-Fano Limit

Pedram Loghmannia, *Student Member, IEEE*, Majid Manteghi, *Senior Member, IEEE*

*Abstract*— In this work, a parametric up-converter amplifier is introduced as a wideband impedance matching network for receiving electrically small antennas. Chu's limit restricts the minimum Q-factor of unloaded small antennas; however, the practical bandwidth of small antennas is defined by their loaded Q-factor. By connecting a small antenna to an amplifier with a real input impedance several times greater than the radiation resistance of the antenna, we propose increasing the return loss to reduce the loaded Q-factor and increase the bandwidth as a result of the Bode-Fano theorem. In addition, a parametric amplifier is used because, in comparison with transistor amplifiers, it offers low noise characteristics. The gain of the low noise parametric amplifier compensates for the loss due to the imposed mismatch. Our simulation result shows bandwidth improvements up to 32 times can be accomplished by trading 2 dB of noise figure compare to 15 dB suggested by Chu's limit for a lossy antenna.

## I. INTRODUCTION

ELECTRICALLY Small Antennas (ESAs) are an inseparable part of a wide range of VHF and UHF transceivers, very low-frequency underwater radios, implanted devices, the Internet of Things, and mobile devices. Demands for higher information rates simultaneously encourage the use of broadband antennas by radio engineers. However, alongside Wheeler and Chu [1, 2], the Bode-Fano bound [3] shows that the realized gain of ESAs (directivity of an ESA is 1.5 for full space and 3 for half-space; therefore, the gain is equal to antenna efficiency times 1.5 or 3 while the realized gain is the gain multiplied by the mismatch factor) needs to be traded for broader bandwidth. Essentially, Chu's limit ties the quality factor of a lossless antenna to the inverse of its electric size to the power of three. This relationship shows how fast the realized gain-bandwidth product is decreased by miniaturization.

Chu's limit was established based on a single-mode and Linear Time-Invariant (LTI) antenna; thus, efforts have been made since the 1950s to crack the limit using multi-mode and/or non-LTI techniques for ESAs [4-6]. Almost all non-LTI techniques, especially for high-power applications, have been implemented for transmit antennas [4]. On the other hand, wireless network engineers had an understanding that part of the loss associated with mismatch and poor radiation efficiency of ESAs could be compensated with low noise amplification on the receiving side (e.g., short-wave and long-wave radio receivers). Nevertheless, utilizing an active component deteriorates the overall noise figure, *NF*, since the amplification happens after the antenna stage. The literature also has suggested many methods to increase an ESA's bandwidth using active matching or non-Foster matching techniques [7-11]. An appropriate figure of merit is to compare the *NF* and the efficiency of any active matching system, respectively, on the receiving and transmitting sides with the simple case of directly connecting the ESA to an amplifier.

In [12], an ESA with a non-Foster matching network was compared with a simple amplifier connected to the same ESA in terms of the noise and gain performance. The front end consists of the ESA connected to an active matching circuit in the receiver (case 1: non-Foster matching network, and case 2: a simple amplifier), as shown in Fig. 1. There is also a preamplifier between the front-end and the spectrum analyzer. The same transistor technology was used in the matching circuit of both cases to have a fair comparison. The noise figure of the preamplifier + spectrum analyzer, $NF_r$, was swept from 0 dB to 16 dB to study both cases' noise performances. The results show that the overall *NF* of the non-Foster receiver chain, case 1, is degraded relative to the overall *NF* of the receiver with a simple transistor amplifier, case 2 when $NF_r$ is less than 6 dB. They also carried out a detailed study of the noise sources [12] in the Linvill Negative Impedance (NIC) converter [13]. They found that the total noise voltage generated in the NIC is proportional to the magnitude of the NIC input impedance. In other words, the larger reactance the NIC (non-Foster technique) cancels, the higher the noise it produces. As a result, it is seen that even compared to the simple amplifier case, the noise performance of the non-Foster technique is not promising.

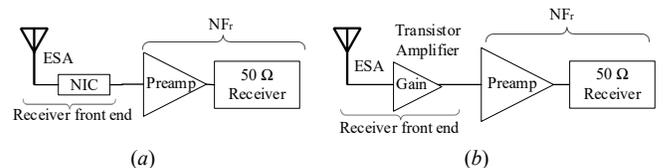

Fig. 1. *a*) case 1: non-Foster technique. *b*) case 2: simple transistor amplifier – direct connection [12].

Due to their applications in the magnetless non-reciprocal devices [14-16], low noise active antennas [17, 18], and wideband impedance matching techniques of small antennas in transmission mode [19, 20], parametric devices as non-linear and/or time-variant components have recently regained the attention of researchers. Although parametric amplifiers are low noise components compared to the transistor counterparts due to their reactive nature, to the best of our knowledge, they have not been used in any ESA matching circuit on the receiving side. This paper introduces an active matching methodology that dramatically increases an ESA's instantaneous bandwidth in the receive side while it does not increase the *NF* at the same rate.

As shown in Fig. 2 (a), the most traditional way of using magnetic (i.e., highly inductive) ESA is connecting it to a matched load leading to an extremely narrow bandwidth. Fig. 2 (b) shows one way to increase the bandwidth by adding loss to the antenna, which results in $(3\times N)$ dB added noise for $2^N$ times bandwidth improvement. Another way, shown in Fig. 2 (c), is removing the tuning capacitor and connecting the antenna directly to a pure real impedance leading to poor noise performance due to a huge mismatch.

We believe, by inspiration of the Bode-Fano limit, tuning the antenna at the desired frequency and connecting it to a pure impedance several times greater than the real part of the antenna input impedance (see Fig. 2 (d)) gives the best trade-off between bandwidth improvement and added noise as explained in section II of this paper. To the authors' best knowledge, the circuit shown in Fig. 2 (d) has never been used in the context of widening the bandwidth of the small antennas. Because the antenna is working on the receive side, $R_L$ in all four circuits shown in Fig. 2 represents the input impedance of an amplifier. In section II, we show that the noise figure of the amplifier is a limiting factor. As a result, we propose using an up-converter parametric amplifier as an extremely low noise amplifier to keep the noise level as low as possible in section III.

The input impedance of the parametric amplifier is tuned to appear as a real impedance to the ESA circuit model and reduce its Q-factor. Remember that this fictitious resistor does not affect the overall noise performance of the system as it does not convert the input power to heat. Furthermore, in section III, we assume ideal circuits for the antenna element and the parametric amplifier to make the discussion simple and straightforward. In section IV, we propose a design for the realized loop antenna. Then, we validate our proposed method for the real structure using hybrid simulations (full-wave analysis + circuit simulator) as detailed in section V. Finally, in section VI, the structure is prototyped, and its received signal is compared to the simulation results.

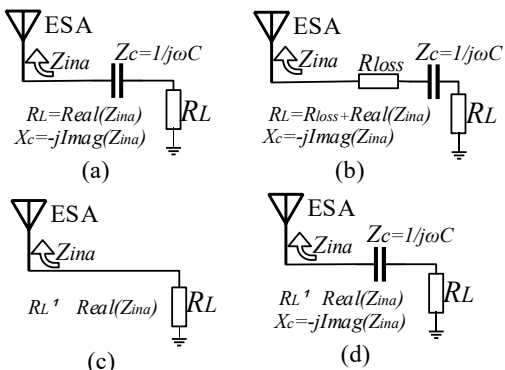

Fig. 2. Different types of loading a magnetic ESA. a) matched case. b) adding loss. c) direct connection, also shown in Fig. 1 (b). d) tuned case using the Bode-Fano theorem (our method).

## II. THE ESAs BACKGROUND AND THEIR BANDWIDTH IMPROVEMENT USING THE BODE-FANO BOUND

In 1948, Chu considered a small magnetic antenna enclosed in a sphere of radius *a* [1], which excites the fundamental spherical mode outside the sphere, $TE_{10}$, but does not store and dissipate any magnetic energy within the sphere. He showed the equivalent circuit of the small magnetic antenna is a series *RC* in parallel with an inductance, *L*. However, the resulting poles in the complex frequency plane are located far above the frequency range that the antenna is still considered small. Furthermore, an ESA has recently been shown to be similar to a lossy inductor (magnetic antenna, e.g., small loop antenna) or lossy capacitor (electric antenna, e.g., a short dipole antenna) [4]. The simplified models of the electrically small magnetic and electric antennas, respectively, are series *RL* and series *RC*. These models are not in contradiction with Chu's model at the frequency range of interest. Electric and magnetic fields are quasi-static in the vicinity of an ESA, and therefore, it stores either electrical or magnetic energy as a first-order circuit [4]. As a result, an ESA's quality factor was demonstrated to be:

$$Q_a = \frac{1}{(ka)^3} \qquad (1)$$

where *k* is the free-space wavenumber.

Investigating Chu's assumptions clarifies that he calculated the antenna unloaded Q-factor. However, we always have to load the antenna with an external circuit (receiver or transmitter) in practical applications. As a result, it is necessary to define the bandwidth of the antenna based on the loaded Q-factor; that is why Chu referred to Fano's paper [3] for the relationship between his Q-factor and the antenna bandwidth.

Fig. 3. (*a*) shows the equivalent circuit (series *RL*) of a receiving magnetic ESA connected to a load, $R_{Lin}$, as the receiver load. $R_a$, $L_a$, and $V_a = \mathbf{E}^i \cdot \mathbf{l}_{eff}$ represent antenna radiation resistance (in addition to the antenna loss), inductance, and open-circuit voltage, respectively. $\mathbf{E}^i$ is the incident electric field at the antenna aperture and $\mathbf{l}_{eff}$ is the effective length of the antenna. Moreover, $C_1$ represents the tuning capacitor in addition to the parasitic capacitor associated with the antenna structure. Increasing the electric size of the magnetic ESA results in a higher parasitic capacitor. This means that the antenna begins to store electric energy; however, stored magnetic energy is still dominant. It is worth mentioning that using Chu's circuit model (instead of using the simplified series *RL* model) leads to a more accurate result, but it makes the equation more complex and tedious. This section uses a simple model to illustrate the concept in a simplified manner, leading to close form relations.

According to Fig. 3. (*a*), the Q-factor of the loaded antenna is defined as:

$$Q_L = \frac{2\pi f L_a}{R_a + R_{Lin}} \qquad (2)$$

where *f* is the center frequency of the tuned antenna. On the other hand, the Q-factor for the unloaded antenna model is obtained by setting $R_{Lin} = 0$ as:

$$Q_a = \frac{2\pi f L_a}{R_a} \tag{3}$$

As seen, it is possible to decrease the loaded Q-factor by increasing $R_{Lin}$ leading to a broader bandwidth. However, simple analysis shows that setting $R_{Lin}$ to any value other than $R_a$ reduces the power delivered to $R_{Lin}$ due to the mismatch at the resonant frequency.

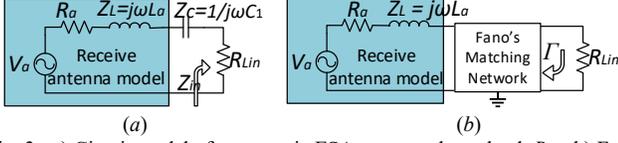

Fig. 3. a) Circuit model of a magnetic ESA connected to a load, $R_{Lin}$. b) Fano's theorem applied to a simple RL circuit.

Fano studied the above concept rigorously and formulated it mathematically [3]. He used the Darlington impedance theorem [21] to conduct a comprehensive study on matching a complex load/source to a pure resistance. His work has resulted in some integral equations defining a trade-off between matching bandwidth and realized gain (efficiency associated with return loss).

Fig. 3. (b) shows a simple series RL circuit as a complex source impedance ($R_a + j\omega L_a$). Fano's simplified integral to match this RL circuit to a purely resistive load $R_{Lin}$ is:

$$\int_0^\infty \ln\frac{1}{|\Gamma(f)|}df \le \frac{R_a}{2L_a} \tag{4}$$

where $\Gamma$ is the reflection coefficient between the matching circuit and the load. Fano assumed that $\Gamma(f)$ = constant value in the entire passband to simplify the integral. Besides, he showed that the right side of (4) for the prescribed $L_a$ and $R_a$ places an upper limit on $\ln(1/|\Gamma|)$ region over the entire frequency band. A small value of $R_a/L_a$ reduces the upper limit, resulting in a reduction of the region under integration (higher $Q_a$ and lower realized gain-bandwidth product).

The critical point is that $R_{Lin}$ does not contribute to the right side of (4) as Fano technically used an ideal transformer to convert the real part of the impedance when needed. This could be misleading in practical application as realizing the transformer could become a more profound new challenge than the original matching problem. In this sense, a high ratio of $R_{Lin}/R_a$, is not a limiting factor in Fano's theorem, while it is a challenge in practical realization. This contradicts [19], which theoretically solved the problem of the high contrast ratio between $R_a$ and $R_{Lin}$ and then claimed to operate beyond the Bode-Fano's bound.

Considering constant value for $\ln(1/|\Gamma|) = \max[\ln(1/|\Gamma|)]$ in the desired frequency range and making it zero at other frequencies as described in [3] leads to the efficient use of the area under the integration and maximizing the bandwidth:

$$\Delta f_{max} = \frac{0.5}{(L_a/R_a)\max[\ln(1/|\Gamma|)]} \tag{5}$$

That requires an infinite number of reactive elements in the matching network. To understand the importance of this equation, we examine a few examples.

Fig. 4. compares the bandwidth of a simple RLC resonator with $R = 50\ \Omega$ and a given Q-factor with the same resonator matched with Fano's matching network for different values of $\Gamma$. Based on the Fano's theorem, the area under the $\ln 1/|\Gamma|$ curve is bounded by $R_a/(2L_a)$ and the wider the frequency bandwidth requires less $\ln 1/|\Gamma|$ which is translated into a higher mismatch. In other words, to increase a resonator's bandwidth with a given Q, a part of the power must reflect back (mismatch). The more reflected power, the broader the bandwidth. Considering $L_a$ and $R_a$ in series as the circuit model of a magnetic ESA, we can use (3) and (5) to link the Bode-Fano to the Chu's Q-factor as:

$$\frac{\Delta f_{max}}{f} = \frac{\pi}{Q_a \times \max[\ln(1/|\Gamma|)]} \tag{6}$$

This equation shows that the maximum bandwidth is proportional to the inverse of Chu's unloaded Q-factor times $\ln 1/|\Gamma|$. Because one can amplify the received signal utilizing an amplifier, the only limitation to increase the bandwidth of a high-Q receiving antenna is the signal to noise ratio ($S_r/N_r$) of the received signal ($S_r/N_r$).

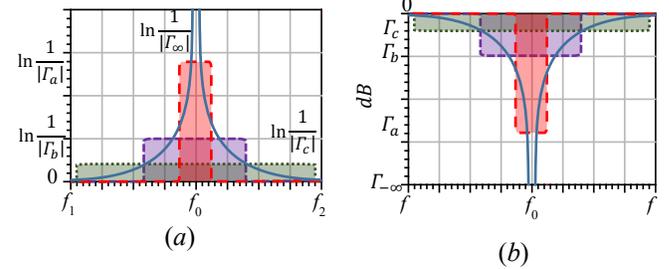

Fig. 4. a) curves with the same area of integration for different values of $\Gamma$. b) return loss for the same $\Gamma's$ in the figure (a) in dB scale. Based on Fano's theorem, the area under the $\ln 1/|\Gamma|$ is bounded by $0.5/(L_a/R_a)$ and the wider the frequency bandwidth requires bigger $|\Gamma|$ which is translated to a higher mismatch.

Based on Poynting's theorem, the available power density at the antenna aperture is $W^i = |\boldsymbol{E}^i|^2/(2\eta)$ and as it was mentioned before, the open-circuit voltage at the antenna port is $V_a = \boldsymbol{E}^i \cdot \boldsymbol{l}_{eff}$ which peaks for matched polarization at $V_a^{\max} = |E^i l_{eff}|$. Thus, the maximum power delivered to $R_{Lin}$ is:

$$\text{Max}(P_{del}) = \frac{\eta}{2R_a}W^i l_{eff}^2(1-|\Gamma|^2) = \frac{1}{4R_a}V_a^2(1-|\Gamma|^2) \tag{7}$$

where $\Gamma$ is the return loss of the Fano's matching network connected to the antenna seen from $R_{Lin}$ (Fig. 3 (b)). The maximum power delivered to $R_{Lin}$ is half of the power generated by $V_a$ (for $\Gamma = 0$) and the antenna backscatters (reradiates) at least half of the power (dissipated power in $R_a$). A higher mismatch (equivalent to wider bandwidth) can increase the backscattered power from the aperture of the receiving antenna from 50% up to 100% (for $|\Gamma| = 1$).

As another example, to show the gain-bandwidth limitation, we have synthesized a 4$^{\text{th}}$ order matching circuit based on the Bode-Fano method [3] for $\max(\ln[1/|\Gamma|]) = 0.348$ and .052 corresponding to 3 dB and 10 dB reduction in the gain, respectively. Results of Advanced Design Simulations (ADS) are provided in Fig. 5 and are compared to both the ideal case (equation (6)) and the simple conjugate matched case corresponding to Fig. 3. (a) with $R_a = R_{Lin}$.

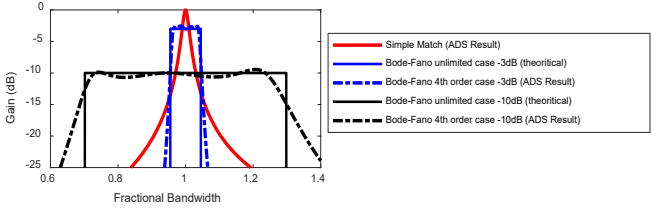

Fig. 5. The Bode-Fano matching method. In all graphs, the unloaded Q-factor is set to 100.

As shown, by reducing the realized gain (higher backscattering by implementing mismatch), one can dramatically increase the bandwidth of a high-Q antenna. Now, let us take $R_{Lin}$ as the input impedance of an active circuit like an amplifier, as shown in Fig. 6. We can set the amplifier gain to 3 dB or 10 dB, respectively, to account for 3 dB or 10 dB losses. In the absence of noise, the antenna bandwidth is improved, and its gain is remedied by an amplifier while still constrained by the Fano's bound (it did not go beyond the Fano's bound, which contradicts the statements in [20]).

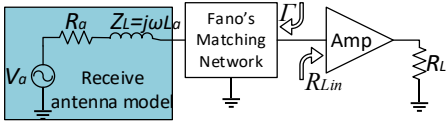

Fig. 6. The Bode-Fano matching method is attached to an amplifier.

However, in reality, utilizing active components degrades system parameters such as *NF*, efficiency, nonlinearity, etc. However, in the vast majority of receiving/transmitting applications, noise/efficiency is the most critical consideration. The *NF* of the front-end of the receiver shown in Fig. 7 is defined as the available signal to noise ratio at the antenna terminal, $S_i/N_i$, divided by the signal to noise ratio delivered to the load, $S_o/N_o$. Assuming the antenna is lossless and its temperature is $T_A = T_0$, one can calculate the noise figure as:

$$NF = \frac{S_i/N_i}{S_o/N_o} = 1 + \frac{NF_{amp} - 1}{1 - |\Gamma|^2} \qquad (8)$$

where $NF_{amp}$ and $T_0$ are the amplifier noise figure and the room temperature (290 K), respectively. In an ideal realization, Fano's matching network comprises only reactive components and does not contribute to the noise. It is lossless, but it is responsible for the realized gain, $G_a$, and contributes to *NF* by reducing the delivered power, as seen in (8). In fact, higher mismatch, $1 - |\Gamma|^2 \to 0$, boosts the effect of the amplifier-generated noise and dramatically deteriorates the overall *NF*. Nonetheless, it is possible to control the mismatch effect with very low noise amplifiers, $NF_{amp} \to 1$. Thus, for the first time, we incorporate the parametric amplifiers as the best candidates for receiving ESA applications to achieve this objective.

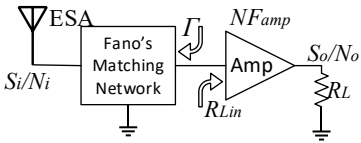

Fig. 7. Noise characteristics of the Bode-Fano matching method.

In contrast to what was mentioned in [20], the parametric impedance matching technique does not apply to the transmitting case due to its low efficiency in comparison to the transistor case or the non-Foster technique. It is also worth mentioning that due to the sinusoidal nature of its power supply, the maximum theoretical efficiency of a parametric amplifier is 50 percent. In this paper, we concentrate only on the receive side. For more details on the transmit side, the reader should refer to [22], which compares the non-Foster technique with the simple amplifier in terms of efficiency on the transmit side.

While Jacob *et al.* [12] confirmed the weak noise performance of the non-Foster technique, equation (8) conceptually verifies that an unfair noise figure comparison has been made to evaluate the noise performance of the non-Foster technique [7]. Fig. 8 (*a*) and (*b*) outlines the noise measurement set-up used in [7] to claim noise performance of the non-Foster technique (see Fig. 8 (*a*)) compared with no matching (direct connection) technique (see Fig. 8 (*b*)). Nonetheless, because of two main reasons, none of the corresponding topologies has an advantageous noise characteristic: First, as seen in Fig. 8 (b), there is a large mismatch, $1 - |\Gamma|^2 \to 0$, between the ESA and the receiver due to the high contrast between $Z_a$ and the input impedance of the receiver (50 Ω). As discussed, one can introduce some mismatch to increase the bandwidth. However, it is necessary to use the Bode-Fano theorem (as shown in Fig. 8 (c)) to achieve the lowest possible mismatch for a specified bandwidth. It should be noted that even the use of a single tuning capacitor or inductor (1$^{st}$ order Bode-Fano circuit) reduces the level of the mismatch for the specified bandwidth and improves the overall noise figure based on (8). Consequently, connecting a High-Q antenna directly to a receiver without matching circuits is not an effective way to increase bandwidth. It should not be considered a reference structure for a fair noise comparison. Second, the effect of the active component noise figure (amplifier in Fig. 7 or receiver in Fig. 8) is essential due to the mismatch. In case b (see Fig. 8 (b)), utilizing a high NF receiver significantly deteriorates the overall noise figure. Therefore, it would be a more appropriate choice to use a low noise amplifier (LNA) such as the one shown in Fig. 8 (c) for a fair comparison of the noise figure. We suggest the circuit shown in Fig. 8 (c) as a good front end to assess any other matching technique's noise performance.

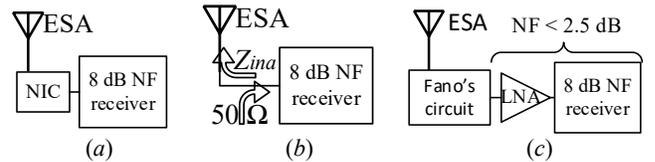

Fig. 8. a) The non-Foster matching technique. b) no matching – direct connection, also shown in Fig. 1 (b). c) The Bode-Fano method: 2.5 dB noise figure is easy to achieve based on the datasheets of the vast majority of commercially available low noise amplifiers.

### III. PRINCIPLES OF THE BROADBAND PARAMETRIC MATCHING TECHNIQUE

In the previous section, we showed that one needs to introduce a mismatch by changing the input impedance of the amplifier to improve the antenna bandwidth. Besides, we saw a low noise amplifier is necessary to keep the noise level low while increasing the bandwidth. This section replaces the amplifier in Fig. 6 with a parametric up-converter amplifier because 1. It offers lower NF compared to its transistor-based counterpart. 2. The input impedance of the up-converter parametric amplifier can be adjusted by changing pump power.

*A. Review of the Up-Converter Parametric Amplifier*

Fig. 9. illustrates the basic topology of the up-converter parametric amplifier consisting of a non-linear capacitor, $C(v)$, (e.g., a varactor diode) in parallel with three filters at the signal ($f_s$), pump ($f_p$), and output ($f_o$) frequencies [23]. Ideally, each filter allows current at the associated frequency and is an open circuit at all other frequencies ($mf_s + nf_p$). Applying a small-signal voltage, $V_a$, at $f_s$ along with a large pumping voltage, $V_p$, at $f_p$ to the non-linear capacitor results in a new waveform at the output frequency ($f_o = f_s + f_p$), which is the amplified version of the input signal, $V_a$. The core element of the parametric amplifier is the non-linear reactive element, $C(v)$. Real power transfers from the pump to the output frequency, and as a result, amplification takes place [24, 25]. In Fig. 9, $R_a$, $R_P$, $R_s$, and $R_L$ are the signal source resistance, pump source resistance, varactor diode loss, and the amplifier load, respectively.

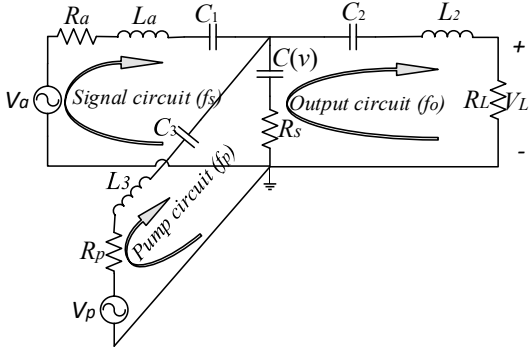

Fig. 9. The basic circuit of the up-converter parametric amplifier. Each loop has zero net reactance at the associated frequency. For example, in the signal circuit loop, $L_a$, $C_1$, and dc-capacitance of $C(v)$ make a resonant circuit at $f_s$.

The non-linear capacitor is modeled as a linear time-variant capacitor using small-signal approximations [23]:

$$C(t) = C_0(1 + 2M \cos(2\pi f_p t)) \quad (9)$$

where $C_0$ and $M$ are the dc-capacitance and the modulation index, respectively, the small-signal model is shown in Fig. 10.

In our work, the series combination of $V_a$, $R_a$, and $L_a$ is considered as the circuit model of a magnetic ESA operating in the receive mode as depicted in Fig. 10. In other words, the received signal is modeled as a voltage source $V_a$ (antenna open circuit voltage) in series with the antenna radiation resistance $R_a$. The sinusoidal waveform at $f_s$ (received by the antenna) is multiplied by the time modulated capacitor at $f_p$ to generate the new harmonic at $f_o = f_p + f_s$. The amplifier conversion gain, $G_a$, is defined as:

$$G_a = \frac{\text{Power delivered to } R_L \text{ at } f_o}{\text{Available power from the source at } f_s} = \frac{4R_a V_L^2}{R_L V_a^2} \quad (10)$$

where $V_L$ is the voltage across the load, $R_L$. Conversion gain was calculated in [23] using (10) at the mid-band. Using the same method, we obtain the conversion gain as a function of frequency:

$$G_a = \frac{4R_a R_L}{\left| \frac{j2\pi f_s (1-M^2) C_0 Z_{TO} Z_{TS}}{M} - \frac{M}{j2\pi f_o (1-M^2) C_0} \right|^2} \quad (11)$$

where,

$$Z_{TO} = R_L + R_S + j2\pi f_o L_2 + \frac{1}{j2\pi f_o C_2} + \frac{1}{j2\pi f_o (1-M^2) C_0}$$

$$Z_{TS} = R_a + R_S + j2\pi f_s L_a + \frac{1}{j2\pi f_s C_1} + \frac{1}{j2\pi f_s (1-M^2) C_0}$$

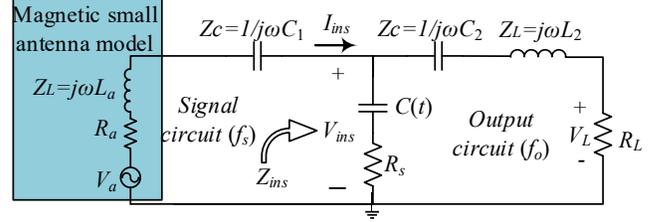

Fig. 10. The small-signal model of the up-converter parametric amplifier.

*B. Wideband Matching Technique*

$f_s$ and $f_o$ must be specified to design the up-converter parametric amplifier. We select $f_s$ = 100 MHz and $f_o$ = 2290 MHz. Design parameters are given in the caption of Fig. 11. The amplifier conversion gain is computed using Matlab and (11) for $R_L = R_a = 20\ \Omega$ (maximum gain condition). The conversion gain is shown with the solid line in Fig. 11. As depicted, the output stage bandwidth (BW= 1.1 MHz) is almost doubled compared to the input stage unloaded resonator bandwidth defined as:

$$\Delta f_S = \frac{f_S}{Q_S} = \frac{R_a}{2\pi L_a} \quad (12)$$

where $Q_S$ is the quality factor of the input stage resonator (Chu's quality factor for the receiving antenna model). The impedance seen by the signal circuit toward the time-variant capacitor at $f_s$ is defined as:

$$Z_{ins} = \frac{V_{ins}(f_s)}{I_{ins}(f_s)} \quad (13)$$

where $V_{ins}$ and $I_{ins}$ are the voltage and current of the time-varying capacitor at $f_s$ as shown in Fig. 10, one can compute the real part of the impedance at the mid-band [23] frequency as:

$$R_{ins} = \text{Re}[Z_{ins}] = R_s + \frac{M^2}{4\pi^2 f_s f_o C_0^2 (1-M^2)^2 (R_L + R_S)} \quad (14)$$

$R_{ins}$ is a positive quantity because $M < 1/2$. As a result, the up-converter parametric amplifier is inherently a stable device compared to the non-foster technique. It is worth mentioning that some other types of parametric amplifiers (e.g., negative resistance or degenerate mode parametric amplifiers) introduce negative resistance to the circuit and might become unstable [23, 24]. $R_{ins}$ loads the input circuit, and the modified (loaded) bandwidth of the input circuit resonator is defined as:

$$\Delta f_{SL} = \frac{f_S}{Q_{SL}} = \frac{(R_a + R_{ins})}{2\pi L_a} \quad (15)$$

where $Q_{SL}$ is the quality factor of the loaded input circuit (receive antenna). For the maximum gain condition (matched case), $R_{ins}$ is equal to $R_a$ resulting in $\Delta f_{SL} = 2\Delta f_S$, which simply explains why the overall signal bandwidth is twice the bandwidth of the unloaded resonator in the input circuit for the maximum gain condition.

According to (14) and (15), reducing the load resistance, $R_L$, increases $R_{ins}$, thus reducing the loaded quality factor of the input circuit and increasing its bandwidth. Fig. 11. demonstrates the amplifier conversion gain calculated using (11) by Matlab for different $R_L$ values ($R_L = R_a$) while other parameters are fixed as given in the caption of Fig. 11. As shown, reducing $R_L$ decreases the conversion gain of the amplifier due to the mismatch between $R_a$ and $R_{ins}$, leading to an increase in the bandwidth as a result of the Bode-Fano theorem. We used the 1st order Bode-Fano circuit to simplify the equations to introduce the mismatch in the input circuit.

Mismatch lowers the received signal level, resulting in the degradation of the overall NF of the receiver. However, the overall NF in the suggested approach is still limited due to two primary factors: 1. the antenna is a part of the amplifier circuit that allows removing the loss of the transmission line between the antenna and the amplifier. 2. due to its reactive nature compared to the transistor amplifier, which uses a trans-resistor part, the parametric amplifier is naturally a low noise device.

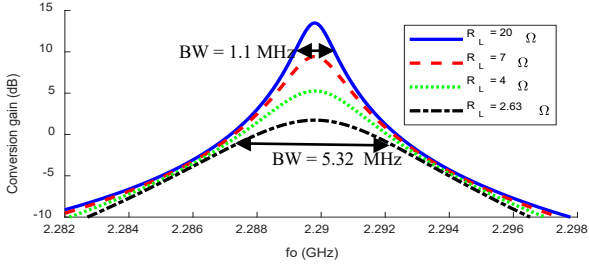

Fig. 11. Amplifier conversion gain versus source resistance (for all graphs $R_L = R_a$, $C_1 = 0.5$ pF, $C_2 = 0.2$ pF, $C_0 = 4$ pF, $M = 0.23$, $R_S = 0.25$ Ω, $L_a = 5.75$ μH, and $L_2 = 25.43$ nH).

We define bandwidth enhancement as the amplifier bandwidth divided by twice the bandwidth of the unloaded resonator in the input circuit ($BW/2\Delta f_s$). Fig. 12. displays the bandwidth enhancement and conversion gain of the amplifier versus $R_L$. As seen, 32 times bandwidth improvement is observed for $R_L = R_a = 2.63$ Ω, corresponding to a total conversion gain of 2 dB (parametric amplifier conversion gain was compensated for the mismatch loss).

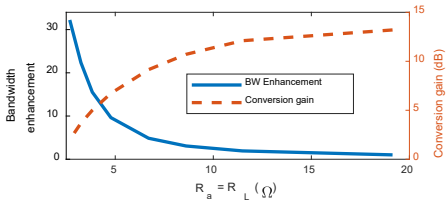

Fig. 12. Amplifier conversion gain and bandwidth enhancement versus source resistance (for all graphs $R_L = R_a$).

### C. Noise Considerations

The up-converter parametric amplifier noise figure is defined as:

$$NF = \frac{N_o}{N_i G_a} = \frac{N_i G_a + N_{amp}}{N_i G_a} \quad (16)$$

where $N_i$ is the input noise power to the amplifier generated by $R_a$ at $f_s$ while $N_{amp}$ is the amplifier noise added due to the noise power generated by $R_s$ at $f_s$ and $f_o$. $R_s$ is the only noise source of the up-converter parametric amplifier. The noise figure of the up-converter parametric amplifier at mid-band frequency has been derived in [23]. We use the same method to formulate the noise figure versus frequency in the entire bandwidth as:

$$NF = \frac{N_s + N_P}{kT_0 G_a} \quad (17)$$

where

$$N_S = \frac{kT_0 G_a (R_s + R_a)}{R_a}$$

$$N_o = \frac{4kT_0 R_s R_L}{\left| Z_{TO} + \frac{M^2}{Z_{TS}[4\pi^2 f_s f_o C_0^2 (1-M^2)^2]} \right|^2}$$

$Z_{TO}$, $Z_{TS}$, and $G_a$ are given in (11). Fig. 13 plots the noise figure for different values of $R_L = R_a$ using Matlab and (17). As depicted, the noise figure is lower than 3 dB in the entire 3dB bandwidth when $R_L = R_a = 2.63$ Ω, corresponding to 32 times bandwidth enhancement.

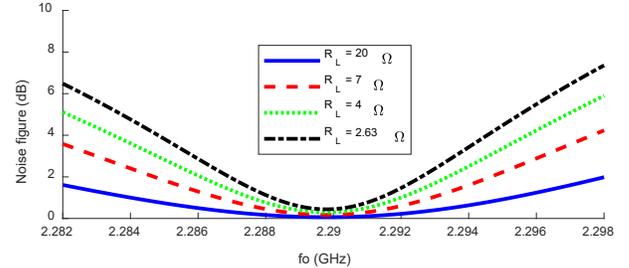

Fig. 13. Noise figure versus source resistance (for all graphs $R_L = R_a$).

### D. ADS Simulations

Fig. 14 shows the ADS harmonic balance (HB) simulation set-up for the circuit model shown in Fig. 9. A series resonant circuit ($RLC$) tuned at $f_p$, and an AC source is used as the pump circuit. This circuit loads the signal and output circuits and slightly detunes their resonant frequencies. $L_a$ and $L_2$ have been reduced by 0.4 and 0.7 percent, respectively, to compensate for the pump circuitry loading effect. All other parameters in the ADS simulations remain unchanged as reported in Fig. 11.

The non-linear capacitor model ($C(v) = C_0+C_{11}v$) in the ADS is used as a core element of the parametric amplifier. $C_0$ and $R_s$ are set to 4 pF, and 0.25 Ω while $C_{11}$ and pump voltage are adjusted to have $M = 0.23$.

Conversion gain and NF are simulated by ADS and are compared with the analytical ones for two different values of $R_L$ in Fig. 15 and 16, respectively. As depicted, ADS results are in close agreement with the analytical ones validating our theoretical analysis.

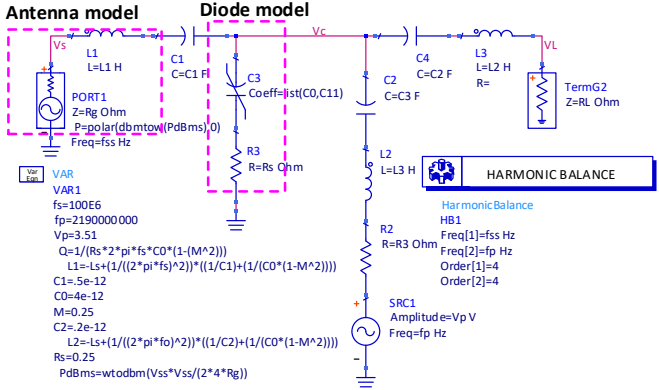

Fig. 14. ADS-HB simulation set-up for the circuit model of the parametric amplifier.

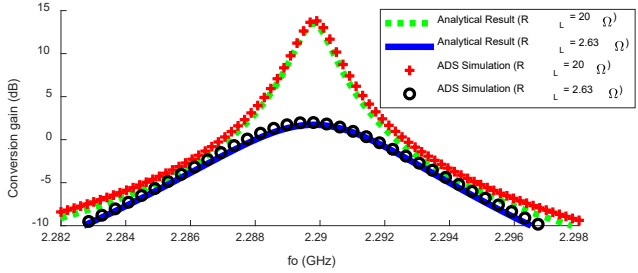

Fig. 15. Amplifier conversion gain for $R_a = R_L = 19.14\ \Omega$ and $2.63\ \Omega$.

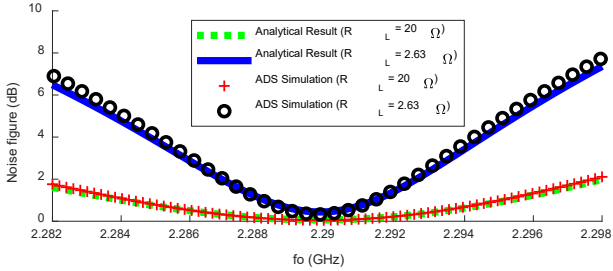

Fig. 16. Amplifier noise figure for $R_a = R_L = 19.14\ \Omega$ and $2.63\ \Omega$.

## IV. Realized Antenna Design and HFSS Simulations

So far, we have considered the primary circuit models for the antenna (see Fig. 6) and parametric amplifier (see Fig. 9). Although circuit models are ideal and do not include parasitic effects (e.g., the loss associated with input and output circuits), they helped us to express the entire structure using the close form relations (see (15) and (17)) leading to gain insight into the concept. In this section, the idea is further developed by proposing a realized parametric amplifier circuit (see Fig. 17), and a realized received antenna model (see Fig. 18). These realized structures are modeled and simulated considering all non-ideal effects, including the parasitic elements.

### A. Circuit Design

In the previous sections, lumped inductors and capacitors have been utilized to implement filters (series $LC$-resonators) at three different frequencies of $f_s$, $f_p$, and $f_o$. In practical designs, the limited Q-factor of lumped components makes them lossy, leading to deteriorating the overall noise figure. Furthermore, it was assumed that each filter is an open circuit at all frequencies other than its resonant frequency, which is hard to accomplish using the basic circuit shown in Fig. 9 and lumped $L$ and $C$ components.

To address these issues, distributed filters and matching circuits are utilized in the design, as depicted in Fig. 17. we briefly discuss the general idea of the design by comparing the basic topology (circuit model shown in Fig. 9) of the parametric amplifier and the proposed design (see Fig. 17) in the following paragraphs.

First of all, $LC$-resonators (filters) at the output ($L_2C_2$) and pump ($L_3C_3$) circuits are replaced with two combline filters offering high rejection while keeping the passband loss minimum. Simulation results show that the input impedance of the combline filter is purely inductive and very small at $f_s$ (<< $f_o$ or $f_p$), making it an almost short circuit at $f_s$. However, as required by the basic topology in Fig. 9, the output and pump circuits are needed to be open circuit at $f_s$. Otherwise, they load the non-linear capacitor at $f_s$ and deteriorate the performance of the amplifier. The varactor diode is placed in series with the signal circuit in the new design to overcome this issue, as depicted in Fig. 17. This way, node $A$ in Fig. 17 is shorted out at $f_s$, and the dc-capacitance of the varactor diode resonates directly with the antenna inductance at $f_s$, leading to removing $C_1$ from the original design. On the other hand, an $\lambda_o/4$ open-circuited stub is incorporated in the signal sub-circuitry to make the node $B$ almost short-circuited at both $f_o$ and $f_p$ frequencies and close the loops of the output and pump circuits. It should be mentioned that $\lambda_o/4$ open-circuited stub behaves like a small capacitor at $f_s$ and has a negligible effect on the performance of the input circuit.

Matching circuit-1 is utilized to match the pump source at $f_p$ and make the pump circuit transparent to the output circuit at $f_o$ by making $Z_p(f_o)$ very large (almost open circuit). Finally, $\lambda/4$-transformer and $L_{out}$ in the output sub-circuitry are incorporated to adjust the real and imaginary parts of the input impedance of the output circuit, $Z_o(f_o)$, to the values required by the up-converter parametric amplifier design.

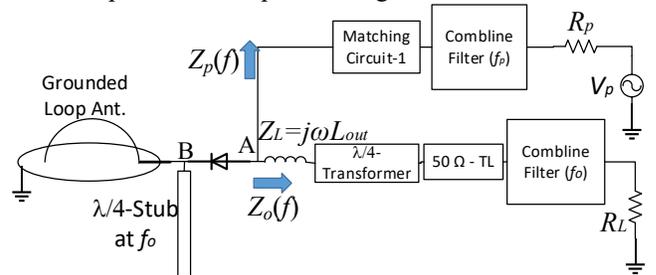

Fig. 17. Block diagram of the proposed design.

### B. Grounded Electrically Small Loop Antenna

A grounded electrically small loop antenna is incorporated as a realized antenna to validate the proposed design. The antenna structure is depicted in Fig. 18. Loop diameter is 18.6 cm ($\approx 0.06\lambda$ at $f_s$), leading to an electrically small antenna.

The antenna is simulated using HFSS, and its input impedance is plotted in Fig. 19, showing the input impedance is highly reactive. Simple $RL$ model (see Fig. 3 (a)) for the grounded loop antenna is obtained by equating the input impedance of the $RL$ circuit model and the antenna input impedance at the exact frequency of 100 MHz. As seen, the simple *RL* circuit does follow the curves associated with both imaginary and real parts of the antenna input impedance, where the frequency approaching the resonant frequency and is not an accurate model. To have a better model, one can consider Chu's circuit for the first mode, $TE_{10}$. However, in the reminding part of the paper, we use the exact model of the antenna computed by HFSS, which shows our proposed technique is valid even for the real antenna since the operational bandwidth is narrow.

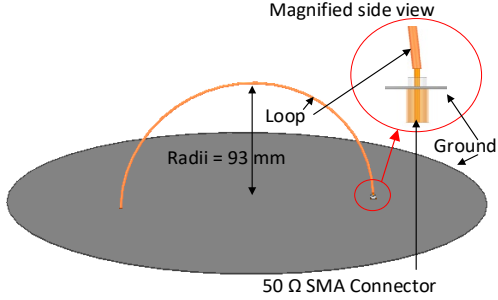

Fig. 18. Grounded electrically small loop antenna.

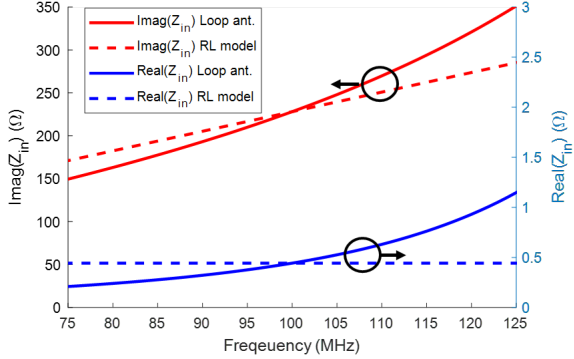

Fig. 19. The input impedance of the grounded loop antenna.

### C. Combline Filters Design and Simulations

Care must be taken into account in the designs of combline filters because they affect the performance of the proposed design in different ways. First, Since the output combline filter defines the loss in the output circuit, its insertion loss directly contributes to the overall noise figure. Thus, the insertion loss of the output combline filter (at $f_o$) should be as low as possible. Second, the pump combline filter should highly reject the noise of the pump circuit at $f_s$ and $f_o$. Otherwise, the noise of the pump source is injected into the output circuit and deteriorated the overall noise figure. Third, although the insertion loss of the pump filter at $f_p$ does not contribute to the overall noise figure, it defines the efficiency of the pump circuit.

$5^{th}$ order combline filters centered at $f_o$ and $f_p$ are designed using the procedure described in [26, 27]. The output combline filter is designed to be wideband (leading to a low insertion loss), while the pump combline filter is designed to be narrow-band resulting in a high rejection at $f_o$ while having a moderate insertion loss. Fig. 20 shows the structure simulated in the HFSS. As depicted in Fig. 21, the passband insertion losses of the output filter centered at $f_o$ = 2.29 GHz and pump filter centered at $f_p$ = 2.19 GHz are 0.3 dB and 1 dB, respectively. On the other hand, the output filter rejection at $f_p$ = 2.19 GHz and pump filter rejection at $f_o$ = 2.29 GHz are simulated to be 40 dB and 80 dB, respectively. As a result, pump to output port isolation is more than 40 dB and 80 dB, respectively at $f_p$ and $f_o$.

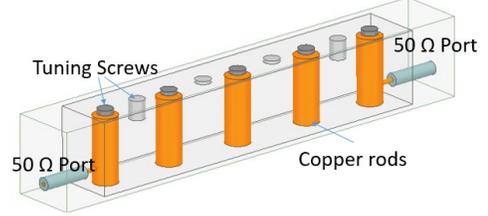

Fig. 20. Combline filter designed at $f_o$.

Fig. 22 illustrates the input impedance of both combline filters around the signal frequency. As shown, the real part is close to zero, and the imaginary part has a small inductive value around 4j Ω at $f_s$ = 100 MHz. It does not alter these properties across the signal frequency by connecting other passive components in the output and pump sub-circuits (including matching network-1, $L_{out}$, and λ/4-transformer) to the combline filters. Then, input impedance seen by the signal circuit toward pump and output circuits at $f_s$ (defined as $(Z_o(f_s) \times Z_p(f_s))/(Z_o(f_s)+Z_p(f_s))$ in Fig. 17) has real part close to zero and small inductive imaginary part. In other words, pump and output circuits load the signal circuit at $f_s$ by small inductive impedance leading to slightly detuning resonant frequency of the signal circuit. The resonant frequency can be tuned again by adjusting the dc-bias of the varactor diode. As a result, point *A* in Fig. 17 can be assumed as a virtual ground (or more accurately as a small inductive load) for the signal sub-circuit at $f_s$. Furthermore, very small inductive input impedance leads to more than 120 dB isolation between the antenna and pump ports (or the antenna and output ports).

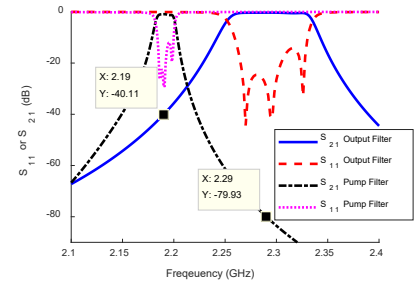

Fig. 21. Scattering parameters of the combline filters around the pump and output frequencies.

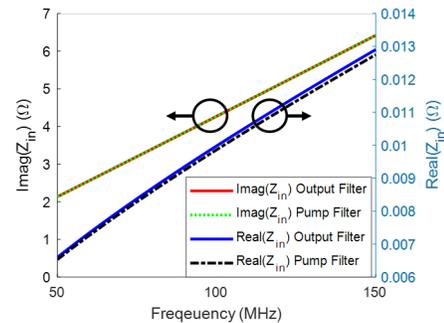

Fig. 22. The input impedance of the combline filters around the signal frequency.

### D. The actual model of the varactor diode

The MA46H2020 varactor diodes from MACOM technology solutions are chosen as the active components in our design.

Two varactor diodes are connected in parallel to lower equivalent series resistance, which is the primary source of noise in the proposed design. The procedure explained in [23, 28] and the non-linear capacitance-voltage characteristics (extracted from the datasheet of MA46H2020) are used to extract $C_0$ and $M$ associated with the time-variant model (see equation (9)) of two diodes in parallel.

Fig. 23 depicts $C_0$ and $M$ versus pump peak voltage for two different DC biases applied to the varactor diodes. An increase in the reverse DC-bias decreases the DC-capacitance, evident from the diode's $C$-$V$ curve. Also, Fig. 23 shows that the modulation index increases by increasing pump voltage.

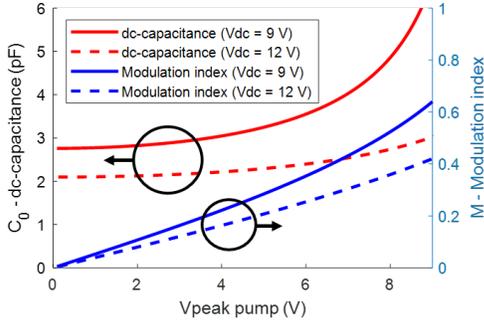

Fig. 23. Time-variant capacitor parameters versus the pump voltage.

It is worth mentioning that data extract from plots like the one in Fig. 23 and concepts explained in III help us evaluate the general behavior of the amplifier and choose the varactor diode.

*E. Feed network*

So far, the design of two combline filters and grounded loop antenna have been explained. Other parts of the proposed design (see Fig. 17) are implemented on a single layer RT/duroid 5880 PCB and is shown in Fig. 24. A small piece of wire in the output circuit is used as the output inductor, $L_{out} \approx 1$ nH. Ports 1, 2, and 3 of the feed network are connected to the antenna, output combline filter, and pump combline filter, respectively. Other parts are implemented as explained in section IV-A.

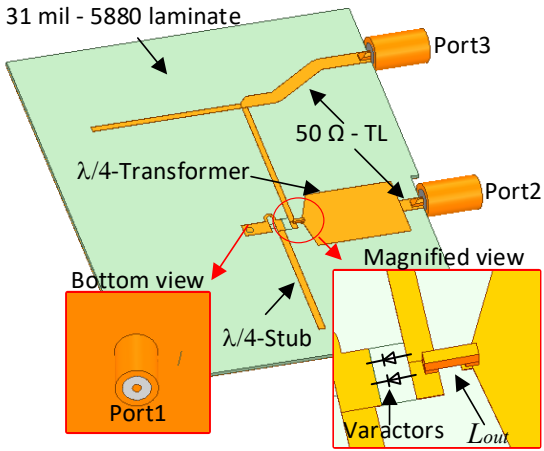

Fig. 24. Feed network structure.

## V. HYBRID SIMULATIONS OF THE REALIZED STRUCTURE

Hybrid simulations using HFSS and ADS are performed on the proposed design to investigate the non-linear behavior. First, as a transmit antenna, a small dipole antenna is placed 1 meter away from the grounded loop antenna, as a receive antenna, and the entire structure (2-ports network) is simulated in the HFSS. Then, the varactor diodes in Fig. 24 are replaced with a lumped port (port4), and HFSS is used to simulate the entire feed network (4-ports network).

The parametric amplifier is a non-linear device that generates all $mf_s + nf_p$ components. All passive components (including combline filters, feed network, and antenna element) need to be simulated in a wide frequency range to model the structure accurately. Accordingly, throughout this paper, all passive components are simulated in the range of 50 MHz to 8 GHz. In addition, It should be remembered that the input and output frequencies of the parametric amplifiers are not the same. For example, in the proposed structure, the antenna is designed to receive electromagnetic waves centered at $f_s$ = 100 MHz while the output power (delivered to the load) is centered at $f_o$ = 2290 MHz. To investigate the bandwidth performance, we sweep the input frequency from 92 MHz to 108 MHz and capture the output power or noise in the frequency range of 2282 MHz to 2298 MHz (the pump waveform is a single-tone sinusoidal at 2190 MHz). As a result, the gain and/or noise of the proposed design may be reported versus $f_o$ or $f_s$ ($= f_o - f_p$). Throughout the paper, the simulated results of the proposed design are plotted versus $f_s$ whenever they need to be compared to the reference structure. Note that both input and output waveforms of the reference structure are centered at $f_s$ = 100 MHz.

In the next step, scattering parameters are imported to the ADS software to perform HB simulations as detailed in Fig. 25. In this set-up, the dipole antenna is fed by a 50 Ω source as a transmitter, and on the receive side, the loop antenna is connected to port one of the feed network. Two varactor diodes in parallel are connected to port 4 of the feed network. As shown, all parasitic elements associated with the package of the varactor are considered in the simulations. Finally, the output and pump combline filters are connected to port2 and port3 of the feed network. It is worth mentioning that the noise and gain results of the ADS simulations include all parasitic effects (including loss associated with the diodes, combline filters, and feed network) because the scattering parameters of the entire lossy physical structure are computed using a full-wave simulation (HFSS).

Another simulation for the simple matched load case has been set up in the ADS to investigate bandwidth improvement of the proposed method. In this simulation, the same receiving loop antenna is directly connected to the matched load, as depicted in Fig. 1 (a). The received power by the matched load is normalized to 0 dB and is depicted in Fig. 26. On the other hand, the received power by 50 Ω load in the proposed method is not normalized to 0 dB, but rather is plotted relative to the received power by the matched load. Although the proposed design receives power at $f_o$ = 2.290 GHz, in Fig. 26, it is plotted versus $f_s$ ($= f_o - f_p$ = 100 MHz) to have both plots in the same figure for the bandwidth comparison. As depicted, the 3 dB bandwidth of the proposed method has been improved 26.5 times compared to the simple matched case. However, for the *RL* circuit model described in section III, we got 32 times bandwidth enhancement. The small difference is due to using the simplified *RL* model in section III, and our simulations show using the exact Chu's circuit model leads to 26.5 times bandwidth enhancement.

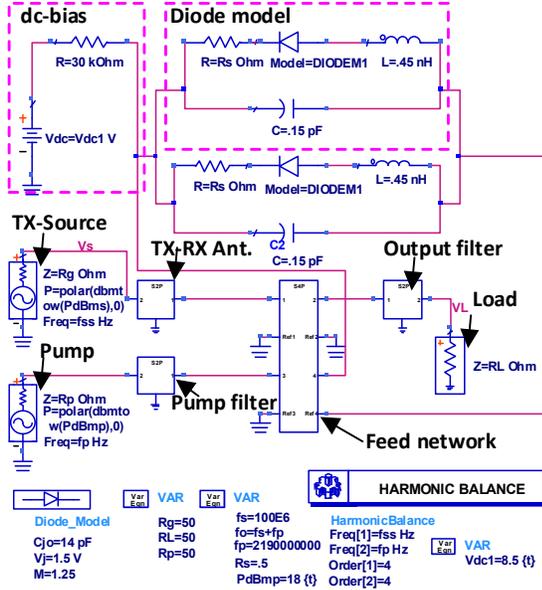

Fig. 25. ADS-HB simulation of the realized structure.

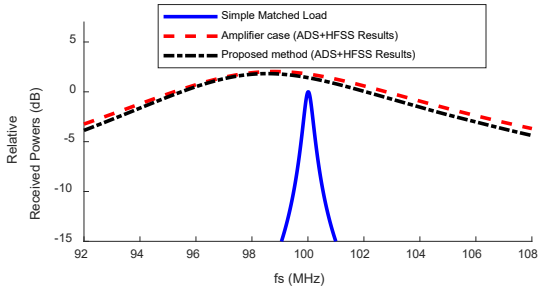

Fig. 26. Relative received powers by the simple matched load case, the proposed method, and the amplifier block. For all cases, the transmit power and the receiving loop antenna are the same. The simple matched load case and the amplifier block case receives power at $f_s$ while the proposed method receives power at $f_o = f_p + f_s$.

Note that according to the circuit shown in Fig. 6, it is possible to increase the bandwidth utilizing a transistor amplifier as well. Therefore, a 1st order Bode-Fano circuit is designed and is connected to an amplifier block in the ADS. The gain and impedance of the amplifier block are defined as 13.5 dB and 50 Ω, respectively. On the other hand, the noise figure of the amplifier block is set to 0.5 dB, 0.75 dB, or 1 dB, which are the typical values for the low noise transistor amplifiers in the market. The mismatch between the antenna and the amplifier block is adjusted to get the bandwidth similar to the proposed design to have a fair noise comparison. Also, the same antenna and the transmit power are used for the amplifier block simulations. The received power (by $R_L = 50$ Ω in Fig. 6) is plotted in Fig. 26. Again, the received power is not normalized to 0 dB and is plotted relative to the received power by the matched load. As shown, the 3dB bandwidth in the case of the amplifier block and the parametric amplifier is almost the same.

The noise simulations for the parametric up-converter and conventional amplifier block are performed in the ADS. In both cases, the effect of the antenna loss in the overall noise figure is ignored because it adds the same value to the overall noise figure of both cases. The overall noise figures, including the effect of mismatch, are shown in Fig. 27. As illustrated, *NF* of the antenna connected to the up-converter active part is 2.9 dB lower than the same antenna connected to an extremely low noise amplifier (e.g., NF = 0.5 dB). It worth mentioning that a lossless capacitor and feeding structure is used to design a 1st order Bode-Fano circuit in the amplifier block case. As a result, the actual noise figure for the realized conventional amplifier block case will be higher. However, the varactor diode used in the parametric amplifier (MA46H2020) is a commercially available component; a lower loss reactive component further decreases the noise figure.

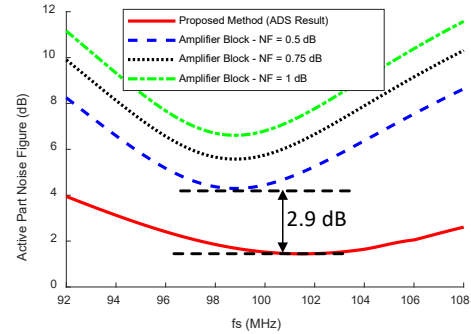

Fig. 27. Overall noise figure comparison. The noise figure for the amplifier block is simulated and plotted at $f_s = 100$ MHz, while the proposed structure is simulated at $f_o = 2.29$ GHz and is plotted versus $f_s = f_o - f_p = 100$ MHz.

The SNR advantage (in dB scale) of the proposed design over the reference structure (the amplifier block – see Fig. 8 (c)) is defined as the received SNR (in dB scale) by the proposed structure minus the received SNR (in dB scale) by the reference structure. The simulated SNR advantage is plotted in Fig. 28 for different values of the amplifier block's NF.

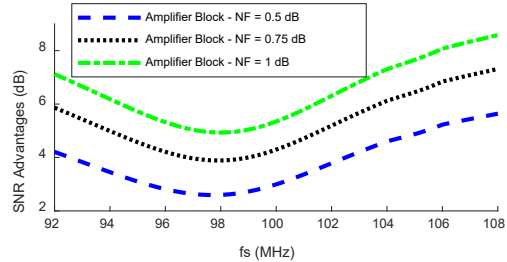

Fig. 28. Simulated SNR advantage. In the proposed structure, the received SNR is simulated at the load and around $f_o = 2.29$ GHz. However, it is mathematically expressed around $f_s = 100$ MHz to compute the SNR advantage over the reference structure.

The pump power and dc bias are set to 18 dBm and 8.5 volts. Considering around 11% efficiency for the pump circuitry (including pump filter loss, synthesizer dc power consumption, etc.), the total power consumption of the proposed structure is 582 mW. The parametric amplifier does not consume any power from the dc source because of zero dc current. Similar ADS simulations have been performed for other methods shown in Fig. 1, and results are compared to the proposed active matching technique in Table I. The TQP3M9036 transistor amplifier from Qorvo is considered as the load ($R_L$ in Fig. 1) for all cases, and its parameters (NF, input P1dB, etc.) are imported in the simulations.

As seen, the proposed technique has the best noise performance, while direct connection has the maximum bandwidth. It should be pointed out that the received power for the direct connection case increases by frequency (due to an increase in the antenna electrical size), and its bandwidth is defined as 3dB variation around 100 MHz. Furthermore, as discussed before, reference [12] confirms that the direct connection case outperforms the non-Foster technique. As shown in Table I, compared to the transistor counterpart, the parametric amplifier needs a higher power to operate, and it offers a lower input P1dB compression point and IIP3.

TABLE I
COMPARISON BETWEEN OUR DESIGN AND OTHER METHODS

| Method \ parameter | BW (MHz) | NF (dB) | Power (mW) | P1dB (dBm) | IIP3 (dBm) |
|---|---|---|---|---|---|
| Proposed design (Fig. 1 (d) + parametric amplifier) | 10 | 1.45 | 582 | -15 | -4.5 |
| Tuned mismatch case (Fig. 1 (d) + TQP3M9036 amplifier) | 10 | 4.3 | 340 | -8 | +2 |
| Adding loss (Fig. 1 (b) + TQP3M9036 amplifier) | 11 | 15.5 | 340 | -8 | +2 |
| Direct connection (Fig. 1 (c) + TQP3M9036 amplifier) | 18 | 13.5 | 340 | -8 | +2 |

## VI. THE PROTOTYPED STRUCTURE AND MEASUREMENTS

The proposed active matching system has been prototyped at the Virginia Tech Antenna Group Laboratory. Fig. 29 shows one of the prototyped combline filters. Aluminum tubes and sheets are used to make the housing, while copper rods are used as resonators. Combline filters of pump and output circuits are tuned using tuning screws. Scattering parameters are measured using R&S ZVA50 VNA and are shown in Fig. 30 along with the simulated results. A small discrepancy between simulated and measured results exist due to handmade prototyping tolerances. However, the critical parameter in our design is the insertion loss, which is acceptable and measured to be 0.55 dB and 1.9 dB, respectively, for the output filter and the pump filter.

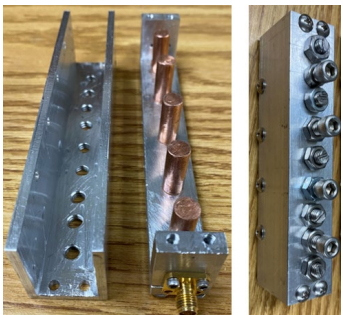

Fig. 29. The prototyped combline filter at the output circuit.

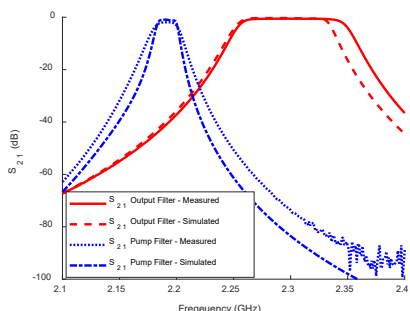

Fig. 30. Frequency responses of the combline filters.

In the next step, the grounded loop antenna and the feed network is prototyped, and all parts are assembled. As shown in Fig. 31, the antenna port is connected to port1 of the feed network, while port2 and 3 of the feed network are connected to the output and pump filters.

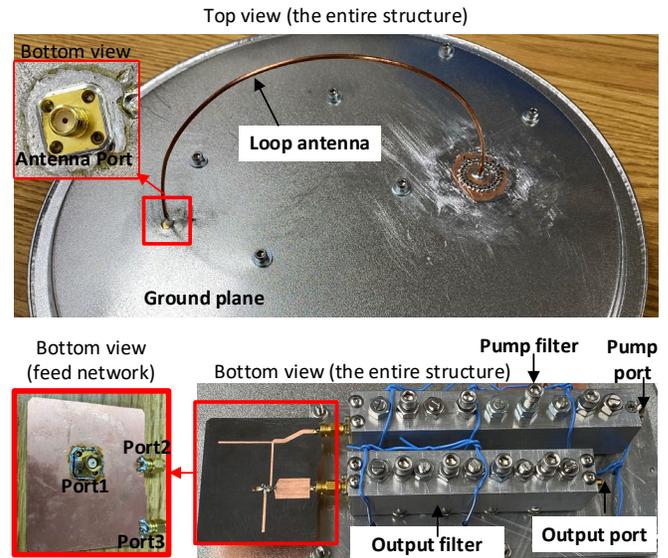

Fig. 31. The prototyped structure.

A measurement set-up, as shown in Fig. 31, is implemented to evaluate the antenna performance. In this set-up, a small monopole antenna is connected to the Tektronix AFG3252 signal generator to form a transmit system. On the other hand, the prototyped structure is configured as a receive antenna, and an R&S FSU spectrum analyzer measures its output power spectrum. The entire structure is placed inside an anechoic chamber, and the distance between the transmit and receive antennas is around 4 meters. Simultaneously, a dc-source and an HP 8648D signal generator are connected to the amplifier as the dc-bias and pump source. The pump source frequency, $f_p$, is set to 2.19 GHz, while the transmitter frequency, $f_s$, sweeps from 92 MHz to 108 MHz. As shown in Fig. 23, increasing the dc-bias of the varactor diode decreases its dc-capacitance, reducing the resonance frequency. This property is utilized to fine-tune the structure leading to the dc-bias of 8.75 V. Finally, the pump power is adjusted to different levels, and the received power spectrums are plotted in Fig. 33. As shown, around 10% (10 MHz) bandwidth is obtained for the pump available power of 138 mW (21.4 dBm). Although the HP 8648D was used as a pump source, in the final structure, the LX2531 synthesizer from Texas Instruments and the QPA9120 driver amplifier from Qorvo can be used to inject 21.4 dBm at $f_p$ to the structure. By adding the power consumption of the LMX2531 synthesizer and QPA9120 amplifier, the overall power consumption of the proposed design is obtained to be 582 mW.

For comparison, a reference set-up is implemented by replacing the proposed structure with a grounded passive loop antenna of the same size as the antenna in our design. The passive loop antenna is matched to 50 Ω and is connected to the spectrum analyzer while the transmitter system and its transmit power stays the same. The bandwidth of the passive antenna is measured to be 0.55 MHz leading to 18 times bandwidth enhancement in the prototyped structure. Note that measured plots in Fig. 33 are relative to the received power by the simple passive matched case. As depicted, the measured receiver gain (parametric amplifier gain + mismatch loss) is 10.2 dB for the matched case corresponding to the maximum gain condition (solid black line in Fig. 33) while it is measured to be 1.25 dB for the 10MHz bandwidth case (solid blue line in Fig. 33).

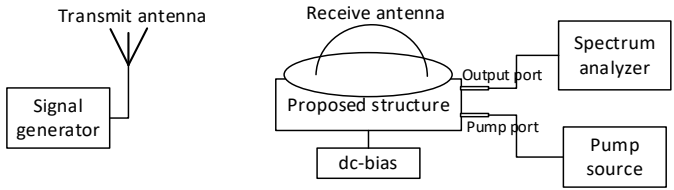

Fig. 32. Test set-up.

Relative simulated results are also included in Fig. 33. As depicted, the same trend exists in both simulated and measured results when we increase the pump power. This trend is explained using the circuit model shown in Fig. 10. For very small values of pump power, $M$ is close to zero (see Fig. 23) and based on (14), the input impedance of the time-variant capacitor, $R_{ins}$, (see Fig. 10) at $f_s$ is significantly smaller than the source resistance, $R_a$, leading to a large mismatch between the source and time-variant capacitor. As a result, the conversion gain is small, and the receiving bandwidth is narrow. Increasing the pump power (increasing $M$), increases $R_{ins}$ and for some specific value of the pump power, $R_{ins}$ becomes equal to $R_a$ leading to the maximum gain condition. In this case, the input and output ports of the parametric amplifier are matched to the source and load, respectively, and the VSWR of input and output ports are close to 1. Further increasing the pump power makes $R_{ins}$ larger than $R_a$, leading to widening the signal bandwidth (see (15)) at the expense of mismatch and gain reduction.

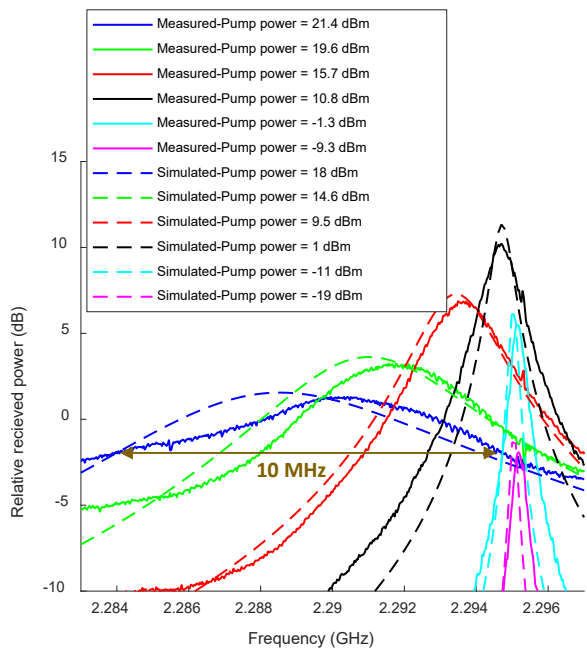

Fig. 33. Relative received powers for different values of the pump power.

Another trend in Fig. 33 is the change in the resonance frequency. According to Fig. 23, increasing the pump power increases the dc-capacitance of the varactor diode leading to a shift down the resonant frequency as depicted in Fig. 33.

As detailed in Table I, the measured P1dB and IIP3 of the proposed structure are -4.5 dBm and -15 dBm, respectively. To measure IIP3, two small monopoles are connected to two sources as the two-tone test transmitter. Another critical parameter in designing a parametric amplifier (same as mixer design) is the isolation between different ports. Table II summarizes the measured port to port isolations. As seen, due to the use of the combline filters, 85.5 dB isolation is obtained between the pump port and the output port at $f_o$, meaning that the noise generated by the pump source is suppressed in the output port. Also, 44.5 dB isolation is achieved between the same ports at $f_p$, leading to suppressing the pump components in the output port.

TABLE II
MEASURED ISOLATION BETWEEN DIFFERENT PORTS

| Frequency | $f_s$ | $f_p$ | $f_o$ |
|---|---|---|---|
| **Pump-Output Isolation (dB)** | - | 42.9 | 85.5 |
| **Antenna-Pump Isolation (dB)** | More than 110 | 44.5 | 98 |
| **Antenna-Output Isolation (dB)** | More than 110 | 61.4 | -24.8 |

Due to the high level of ambient noise in the vicinity of $f_s$, we cannot measure and validate the noise performance of the prototyped structure. However, we rely on the simulated results (see Fig. 27) to validate the noise performance of the design. Furthermore, Fig. 33 confirms the fair agreements exist between the simulated and measured received powers for different situations (e.g., pump power level). As a result, the noise performance of the prototyped structure should not depart from the simulated results.

## VII. CONCLUSION

A review of the Bode-Fano bound and Chu's limits were given in the context of the electrically small antenna as an *RLC* resonator. Then, a low noise wideband impedance matching technique was introduced for ESAs. First, theoretical outcomes were compared to the circuit model simulations. Then, the circuit model was converted to a distributed model, and full-wave simulations of the realized structure were conducted. The overall noise figure of the structure was compared to the transistor case to show the effectiveness of the proposed method. Incorporating a higher-order Bode-Fano circuit would improve the noise figure in both transistor and up-converter cases. However, in this paper, we used the 1st order Bode-Fano circuit to simplify the problem and conceptually explain the effectiveness of the method. Finally, the structure was prototyped, and measured received signals were compared to the simulated ones. The simulated and measured results showed a significant improvement in the noise performance of our method compared to the techniques shown in Fig. 2 (a) to (c).